# Application of Bragg superlattice filters in low temperature microrefrigerators

Gurgen G. Melkonyan[*], Armen M. Gulian[+] and Helmut Kröger[*]

[*]Département de Physique, Université Laval, Québec, Québec, Canada, G1K 7P4

[+]USRA/US Naval Research Laboratory, Washington, DC 20375, USA

## Abstract

We propose to use the Bragg interference filter technology for fabrication of microrefrigerators. The idea of using superconductor-insulator-superconductor (SIS) or normal metal–insulator–superconductor (SIN) tunnel junctions as cooling elements in micro-refrigerators is attractive because of the absence of (micro-) refrigerators operating below 150 K. There are corresponding experiments [1] on SIN tunnel junctions where an attached to the SIN tunnel junction membrane was cooled down. Theoretical approaches (both phenomenological [1] as well as microscopic [2] show that the cooling effect exists also in SIS tunnel junction. However this was not observed experimentally because of inefficient thermal contact between SIS tunnel junction and the membrane that must be cooled. The microscopic approach to cooling is based on the "phonon deficit effect" [2] in nonequilibrium regime of tunnel junctions. In some circumstances, when the applied voltage does not exceed the superconducting energy gap ($\Delta$) the probability of phonon absorption from the heat-bath is higher than its emission in the nonequilibrium regime of tunnel junctions. There is an appropriate absorption window in the phonon emission spectra [2,3] and by absorbing these phonons from the heat bath the SIS or SIN tunnel junction can refrigerate its environment. This effect can be improved by use of phonon filters placed between the tunnel junction and the heath-bath [4]. Such a filter can be the Bragg interference superlattice (Bragg's grating) which is well studied for problems of optical communications. Bragg interference filters are used also for detection of phonons emitted by tunnel junctions [5]. Usually such filters cut low and high frequencies, and the used detector may detect well separated frequencies. In contrast, to enhance the refrigeration process one needs filters with very broad spectral transmission properties or a large transmission band with one or two narrow stop bands. The type of the needed filter will depend on the types of the used tunnel junction. Corresponding discussion is presented.

## 1. Introduction

The absence of the solid state microrefrigerators (SSM) below 150 K stimulates the search for the new electronic cooling devices operating below this temperature. The solid state microrefrigerators have many advantages over other cooling devices: they are long lived, compact and simple at exploitation. The known SSM utilize the thermoelectric effect in semiconductor based divices. For example, the flow of electric current in the semiconductor−semiconductor contacts is accompanied by the heat transfer from the semiconductor into the other semiconductor (this is the well known Peltier effect). Thermoelectric refrigerators are very promising devices for achieving cooling in a completely solid state design. The operational principles of thermoelectric refrigerators are very elegant and described in textbooks (see, e.g., Ref. 6). Thus we will not discuss them in detail to distinguish the different versions available. Rather, we will discuss general principles that disallow the application of these devices at low temperatures. In brief, the functional effectiveness of any thermoelectric device– a generator converting heat into electrical energy or a refrigerator, converting electrical energy into the removal of heat, is governed by the dimensionless figure of merit $ZT$ :

$$ZT = \frac{S^2 \sigma}{k} T. \tag{1}$$

In this expression $Z$ is the (dimensional) figure of merit of thermoelectric material, $T$ is the temperature at the given state, $S$ is the absolute thermoelectric power (so called Seebeck coefficient), and $\sigma$ and $k$ are the electric and heat conductivity respectively. The "golden rule" in the theory of thermoelectric devices is that for the effective operation, the value of $ZT$ should be not much smaller than 1. More precisely, $ZT = \infty$ corresponds to Carnot efficiency, $ZT = 1$ to the efficiency 10%, etc. At some point, when $ZT$ is too small, the heating of the device by the external heat inflow is more than the cooling, so the devices become ineffective as a refrigerator. The universality of the $ZT$−criterion simplifies the consideration of applicability of any thermoelectric device and reduces the problem to the analysis of figure of merit of its constituents.

Considering practical restrictions implied by Eq. (1), one should bear in mind the temperature dependence of $S(T)$, $\sigma(T)$ and $k(T)$. Another important point is that while only electron system contributes into the value of $\sigma(T)$, many quasiparticle systems, especially phonons, can contribute into the heat conductance $k = k^{el} + k^{ph} + ...$. At room temperature, the Wiedemann−Franz law is typically obeyed, and $k/\sigma = k^{el}/\sigma = L_0 T$, where $L_0$ is the Lorenz number ($\cong 25 nW\Omega/K^2$). Then $ZT \approx S^2/L_0 \approx 1$ implies $S \approx 160 \mu V/K$. This value of the Seebeck coefficient can be readily met at room temperatures. Given the convenience of thermoelectric cooling, it is thus not surprising that effective thermoelectric refrigerators exist in the range of $T = 300 \pm 150 K$.

At lower temperature the difficulties start. The problem is not only in general tendency of $S(T)$ to decrease with the temperature ($S(0) = 0$). Rather, the phonon system contributes much more to the heat transfer in solids. About the approximate temperatures $T \approx \hbar \omega_D /(10 k_B)$ ($\omega_D$ is the Debye frequency, $k_B$ is the Boltzmann constant), the value of $ZT$ catastrophically decreases, which effectively rules out application of thermoelectric devices near the liquid nitrogen temperatures. In order to go round these problems one should search for new materials or new devices operating below the temperatures where the traditional coolers fail to operate. Such a device can be the "phonon deficit effect" microrefrigerator discussed below.

This paper is organized as follows. First in the Section 2 we present the ideas which brings to the "phonon deficit effect". A possible superconductor microrefrigerator design is presented in the Section 3. In the Section 4 we discuss our calculated filters for this microrefrigerator.

## 2. Nonequilibrium superconductivity and "phonon deficit effect"

The "phonon deficit effect" (PDE) has been studied sufficiently (see, e.g., [7]) for its essence to be clear. If a superconducting film is immersed into a heat−bath and its electron system is shifted into an excited nonequilibrium state by an external supply of energy, then under appropriate conditions the film absorbs phonons from the heat−bath (rather than emits them to it) in some spectral interval of phonon frequencies. The PDE was initially predicted [2] as an effect accompanying the enhancement of superconductivity [8] in a high−frequency electromagnetic field (UHF). Later it was recognized [9] that the PDE has a more general nature and could characterize situations related to the violation of detailed balance under the influence of external fields. To understand this comment, consider a specific case of a superconductor in the UHF field.

In superconductors at temperatures $T \neq 0$ there is always a certain number of electronic excitations above the energy gap $\Delta$. These "quasiparticle" excitations are in thermodynamic equilibrium with the phonons. Those phonons in superconductors that have energy $\hbar\omega_q$ that is more than the twice of the gap value ($\hbar\omega_q > 2\Delta$) effectively recycle the energy: they pair–break to create quasiparticles then recombine and emit the phonons of the same energy. According to the principle of detailed balance, the probabilities of the direct and reverse processes are identical. The situation changes when an external high–frequency electromagnetic field is applied. If the frequency of the field does not exceed the quasiparticle creation threshold ($2\Delta/\hbar$), the high–frequency electromagnetic field merely changes the "center of gravity" of the quasiparticle distribution function toward the higher energies. The number of excitations left at energies immediately above the gap edge falls below its thermodynamic equilibrium value. Then the probability of phonon absorption at frequencies $\hbar\omega_q \geq 2\Delta$ becomes greater than the probability of their emission. This indicates the violation of the detailed balance in the electron–phonon interaction in presence of an external field. As a result, in a frequency range near $\hbar\omega_q \approx 2\Delta$ there is a deficit of phonons in such a nonequilibrium superconductor. This deficit must be compensated from the outside of the system if the entire picture is to remain stationary. Thus, negative phonon fluxes arise which can be regarded as the phonon flow from the heat–bath to the superconductor.

The situation is essentially similar in case of thin superconducting tunnel junctions. The case of symmetric junctions was considered in detail in [9]. In the sub–threshold tunneling regime ($V < 2\Delta/e$) the result is shown in Fig. 1. An analogy between this case and the superconductor in the UHF strikingly demonstrates the universality of the PDE mechanism.

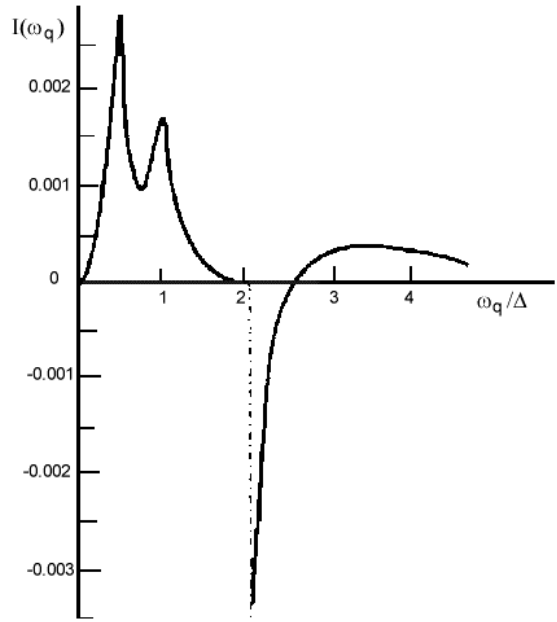

**Fig.1.** The spectral dependence of phonon emission from SIS–junction in sub–threshold regime (eV<2$\Delta$) at T=0.2$\Delta$, eV=0.5$\Delta$.

The question arises whether the PDE could be used to achieve the microrefrigerator device. To answer this question one should calculate the value of total energy, which the nonequilibrium phonons deliver into the heat−bath. This is equal to the integral of the spectral dependencies, such as is plotted in Fig. 1, taken with the weight factor, $\propto \omega^3$, which strongly enhances the relative contribution of the positive "tail". The chance that this integral has a net negative value is higher in the case of asymmetric junction SIS'. Figure 2 plots the nonequilibrium phonon flux after multiplication by $\omega^3$ from the smaller gap superconductor, $\Delta_S \ll \Delta_{S'}$.

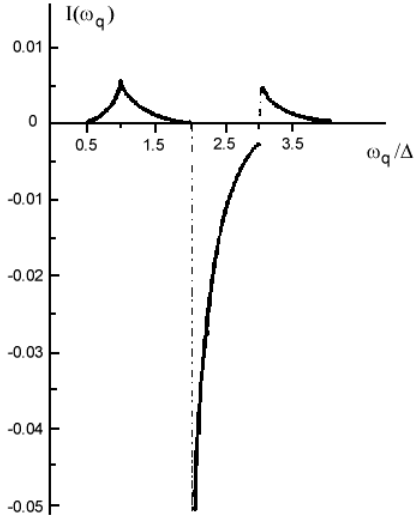

**Fig.2.** Spectral dependence of the phonon flux (in units of energy per unit time) irradiated by the S−film (with the smaller gap $\Delta^{(S)}$) of SIS'−junction.

As follows from the Fig. 2, the net energy−integrated flux is negative [2]. At the same time, there may be sufficient Joule heating in the barrier for the total heat−flux to be positive. Therefore, the application of the idea to refrigerator devices needs a special consideration.

3. Superconducting Microrefrigerator

To enhance further the consequences of the PDE and achieve desirable refrigerator performance, one should more effectively produce a spectral separation of negative and positive fluxes across an interface of nonequilibrium superconductor and the attached heat reservoirs. Suppose, that there is a phonon spectral filter placed in between superconductor S and Reservoir 2 (Fig. 3). Let the filter be transparent in the phonon frequency region where the flux in Fig. 2 is negative. This filter will permit the flow of corresponding phonons ($\hbar\omega_q \approx 2\Delta$) from the Reservoir 2 into the superconductor S. At the same time the filter should block the lower and higher energy phonons, including those created by the heated barrier. Yet another filter with different (reciprocal) transmission properties may be arranged in between the Reservoir 1 and the film S': it will prevent phonons of the energy $\hbar\omega_q \approx 2\Delta$ from being absorbed in S from the side of S' (i.e., from the Reservoir 1), so that the absorption from the Reservoir 2 will be performed more effectively. Then the system shown in Fig.3 will act as an effective refrigerator, able to cool down the temperature of the Reservoir 2. This cooling cannot reduce the temperature much below the critical temperature of superconductor S (Fig.3). But there are different classes of superconductors with critical temperatures covering the wide range of temperatures from about 150K to very low temperatures. Then, cooling cascades can be organized in a manner that is used to be effective for thermoelectric refrigerators. These designs for the PDE coolers deserve careful consideration for practical usability.

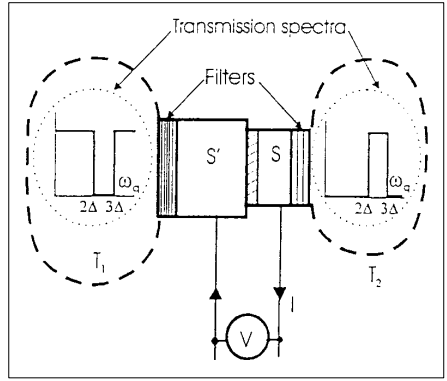

**Fig.3**. Layout of the microrefrigerator with the improved properties. Insets show suggested spectral characteristics of the superlattice filters

## 4. Phonon filters

From the analysis in the Section 3 follows that designing of appropriate phonon filters is of primary importance to the PDE refrigerators. Narayanamurti et al. [5] using the analogy between reflection of plane electromagnetic wave at the interface between two optical media with different refractive indexes $n_1$ and $n_2$ and the acoustic wave reflecting at the boundary of two elastic media with the different acoustic impedance $Z_1$ and $Z_2$ have developed appropriate phonon filter design for their phonon spectrometer. Methods of calculation of the transmission and reflection coefficients in a linear media are described also in an impendance language for the plane acoustic waves [10] and four terminal network theory in general [11], as well as transfer matrix formalism [12].

Here we consider the simplest case of phonon filters for the microrefrigeration processes when acoustic phonons propagate through a superlattice perpendicular to the interfaces. In Fig. 4 the supperlattice system under consideration is presented. The system consists of a repetition of altering layers of material 1 with thickness $d_1$ and impedance $Z_1$, material 2 with thickness $d_2$ and impedance $Z_2$ and so on. If one repeats this configuration $p$ times then superlattice will have a period $D = d_1 + d_2 ... + d_N$ and $L = Np$ layers. The interfaces are perpendicular to the wave vector of incident wave. The supperlattice is placed between two media with the acoustic impendance $Z_{med}$ (media) and $Z_{sub}$ (substrate). It can be shown [11,13] that the fields at the first interface of the superlattice are given in terms of the fields at the last interface of the supperlattice by the matrix equation

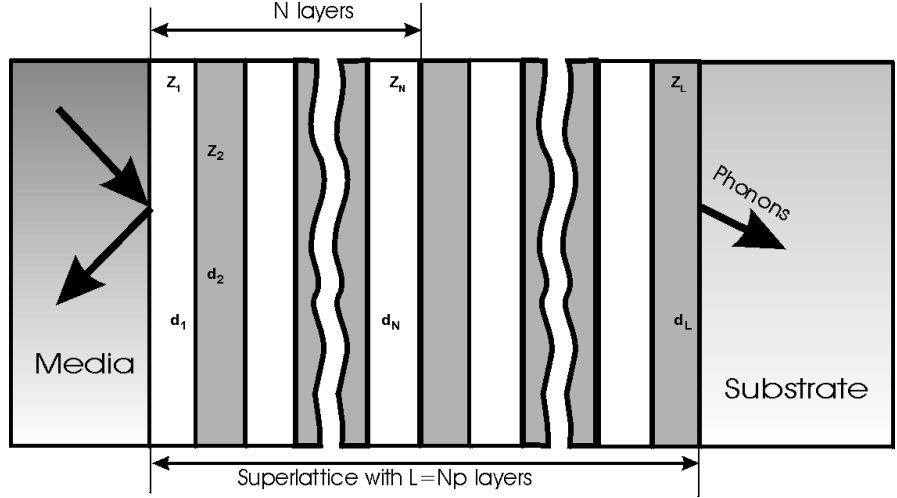

**Fig. 4** Bragg multilayer between two media.

$$\mathbf{U}_1 = \mathbf{M}\mathbf{U}_2 \,. \tag{2}$$

Here $\mathbf{U}_1$ and $\mathbf{U}_2$ are column vectors. $\mathbf{M}$ is the characteristic matrix of the supperlattice or a four terminal network [11]. Expressing $\mathbf{M}$ in terms of the individual layers one can calculate the net transmission or reflection spectra of the acoustic phonons at the interface supperlattice–substrate. In our analysis we follow Narayanamurti et al. [5], and the acoustic properties of each layer is expressed in terms of a characteristic matrix $\mathbf{M}_i$, where $i$ numerates the layers. Then the properties of the supperlatice is described as an equivalent layer with the matrix $\mathbf{M}$ in the equation (2) which is a product of the characteristic matrices of the individual layers

$$\mathbf{M} = \mathbf{M}_1 \mathbf{M}_2 .... \mathbf{M}_L$$

from which the transmission and reflection coefficients can be derived. Hear $L$ is the total number of the layers. The form of characteristic matrices of the multilayers can be found in Refs. 11 and 13.

    To know the needed phonon filter parameters now we will analyze the emission spectra of a nonequilibrium tunnel contact. The spectrum which is presented in Fig.2 is a typical representative for the asymmetric superconductor–other superconductor nonequilibrium tunnel junctions. First we consider the properties of the filter between Reservoir 1 and superconductor S'. This filter should have a broad stop band between frequencies $\hbar\omega_q = 2\Delta$ and $\hbar\omega_q = 3\Delta$ with the relative bandwidth $\hbar(\omega_{q_1} - \omega_{q_2}) \approx 0.4\Delta$. Because of importance of all phonon frequencies in the PDE based microrefrigerators, this filter should work equally well as an ideal stop band filter for some frequencies and as an ideal band pass filter for other frequencies just above and below the stop band. The second filter between Reservoir 2 and S has more complex characteristics, one pass band with bandwidth equal to the stop band of the other filter which is placed between Reservoir 1 and superconductor S' and two stop bands one of which at low frequencies has more broad bandwidth than the pass band of this filter. In brief one needs the reciprocal behavior, i.e., the resonant transmission of the phonon flux in the given frequency range, and the opacity for other frequencies for both filters. Results of our calculation are presented in Figs. 5 and 6. We should notice that all filters have an oscillating behavior of the transmission coefficient in the pass band. This is typical

for the band pass filters. To this question we will return later in this Section. In Table 1 we list the impendance and thickness of the supperlattices used in our calculations.

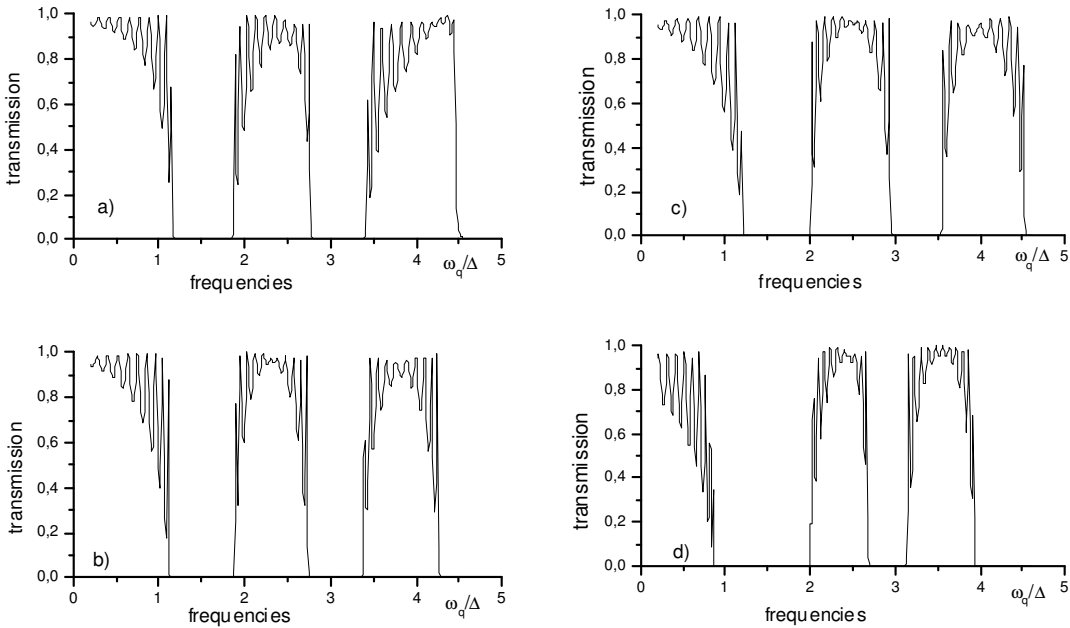

**Fig. 5** Designed band pass filter properties (for the parameters see Table 1).

In Fig. 5 are presented four curves for the transmission spectra of the filter between Reservoir 1 and superconductor S with the different acoustic impendance, layer thickness and layer number. For simplicity the impendance of both the substrate and the media were kept constant. As a media material we have chose copper (it is not a superconducting material) and *Si* as the substrate material. From the comparison of the Fig. 5 a) and Fig. 5 b) we see a gain for the net reflection at low frequencies, with a little lost in the transmittance at intermediate frequencies. Referring to Fig. 2 we see a big phonon fluxes at frequencies about $\hbar\omega_q = \Delta$ and $\hbar\omega_q = 2\Delta$. That filter in Fig. 5 d) may work well for these frequencies. Fig. 5 c) shows that a filter with this transmission band can completely cut out high frequencies and pass over intermediate frequencies with the transmittance practically equal to 1 for these phonons. Our analysis shows that a small variation of the parameters (see Table 1) of the phonon filters will not affect dramatically on the transmission or the reflection properties of these filters.

Figure 6 represents the filter between Reservoir 1 and superconductor S'. From the Table 1 one can note that this type of filters is simple in design. They do not have many periods in contrast to the filter for the other end of the superconductor−superconductor junction. Curves in the Fig. 6 show that the number of periods is not a crucial parameter for this type of filters.

In summary, we have shown that despite of the complexity of the phonon emission spectra of the nonequilibrium SIS' tunnel junctions, the Brag superlattice idea can be used to design appropriate phonon filters for the PDE based microrefrigerators. Here we have not used any modulated parameter to control the oscillations in the pass band. Meanwhile, as pointed out in Ref. 14 the modulation of one of the superlattice parameters can improve the transparency of pass bands, without changing their positions. Also we should mention that applying the refining methods [15] one can improve the transmission properties of these filters. Recent progress in superlattice nano−engineering inspires a hope that the appropriate filters can be constructed once they have been designed theoretically [16,17].

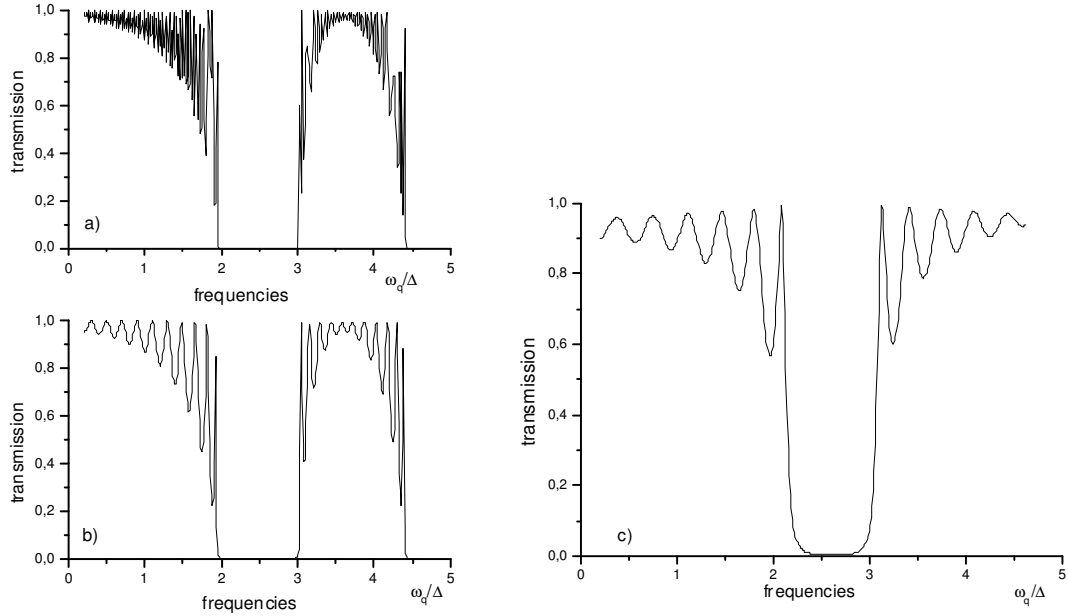

**Fig. 6**. Designed stop band filters (for the parameters see Table 1).

**Table 1.** Parameters of the presented filters (see Figs. 5 and 6). The impendence (Z) and the sizes of the layers (d) are given in units of $10^5$ g/(s cm$^2$) and in angstroms (Å) respectively.

|       | $Z_1$ | $Z_2$ | $Z_3$ | $Z_4$ | $Z_5$ | $d_1$ | $d_2$ | $d_3$ | $d_4$ | $d_5$ | N | p |
|-------|-------|-------|-------|-------|-------|-------|-------|-------|-------|-------|---|---|
| Fig. 5 |       |       |       |       |       |       |       |       |       |       |   |   |
| a     | 2.5   | 1.8   | 1.4   | 0.9   | 1.8   | 50    | 50    | 50    | 50    | 30    | 5 | 11 |
| b     | 3     | 1.8   | 1.3   | 0.9   | 1     | 50    | 50    | 50    | 50    | 30    | 5 | 11 |
| c     | 3     | 1.8   | 1.3   | 0.9   | 1     | 40    | 45    | 50    | 50    | 30    | 5 | 11 |
| d     | 3     | 1.8   | 1.3   | 0.9   | 1     | 40    | 40    | 40    | 45    | 30    | 5 | 11 |
| Fig. 6 |       |       |       |       |       |       |       |       |       |       |   |   |
| a     | 3     | 1.8   | 1.4   |       |       | 43    | 45    | 43    |       |       | 3 | 121 |
| b     | 3     | 1.8   | 1.4   |       |       | 43    | 45    | 43    |       |       | 3 | 11 |
| c     | 3     | 1.8   |       |       |       | 90    | 50    |       |       |       | 2 | 7 |

## 5. Acknowledgment

This work has been supported by NSERC Canada. G. M. would like to thank the "Programme québécois de bourses d'excellence" of Québec for the financial support.## References

[1] M. M. Leivo, J.P. Pekola and D.V. Averin, Appl. Phys. Lett. **68** (1996) 1996; A.J. Manninen, M.M. Leivo and J.P. Pekola, Appl. Phys. Lett. **70** (1997) 1885.8